\documentclass{aa}

\usepackage{mathptm} 
\usepackage{epsfig} 

\begin{document}

\title{Upper Limits on the Pulsed Radio Emission from the Geminga Pulsar 
at 35 \& 327 MHz}

\author{R. Ramachandran\inst{1,2} \and A. A. Deshpande\inst{1} \and C.
        Indrani\inst{1} }

\offprints{R. Ramachandran}

\institute{ Raman Research Institute, C.V. Raman Avenue, 
             Bangalore -- 560080 India: 
             desh@rri.ernet.in, indrani@rri.ernet.in
         \and
              Astronomical Institute, ``Anton Pannekoek'', University
              of Amsterdam, Kruislaan 403, NL-1098 SJ Amsterdam, The
              Netherlands. email: ramach@astro.uva.nl }
\authorrunning{Ramachandran et al.}
\titlerunning{Radio emission from Geminga}
\thesaurus{06(08.16.7) J0633+1746}
\date{Received ; accepted }
\maketitle

\begin{abstract} 
We report here our observations at 35 MHz and 327 MHz made in the direction 
of the $\gamma$--Ray pulsar Geminga. Based on the observed absence of any
significant pulsed emission from this source above our detection thresholds
at the two frequencies, we obtain useful upper-limits for the average 
flux to be 75--100 mJy at 35 MHz, and 0.2--0.3 mJy at 327 MHz. We discuss a 
few possible reasons for the ``radio-quiet'' nature of this pulsar at 
frequencies other than around 100 MHz.
\end{abstract}

\keywords{pulsars: individual: J0633+1746}

\section{Introduction}
\label{sec-intro}
Geminga, as known today, the $\gamma$--ray source in the Gemini constellation, 
was first observed by NASA's SAS--2 $\gamma$--ray astronomy mission (Fichtel
et al. 1975). An enhancement of a little more than a hundred photons, 
in the anticentre region was detected and was thought to be from
an extended nebula, since the initial attempts to associate it with any
known source were unsuccessful. Noting the absence of any radio 
pulsar close to $\gamma 195+5$ (as the source was then called), the papers
announcing the discovery (Kniffen et al. 1975; Fichtel et al.
1975) made an explicit mention of the IC443 SNR close to the source. 

The first systematic radio observations in the direction of the $\gamma$-ray
source, carried out by Bignami et al. (1977) at 610 MHz, did not result in a detection.
With the shrinking error bars on the position of the $\gamma$-ray source, 
it soon became clear that the source is more
like a point source (Masnou et al. 1981), and this led to several
attempts to search for radio emission, in particular, to search for
a radio pulsar (Mandolesi et al. 1978; Mayer-Hasserlwander 
et al. 1979; Seiradakis 1981; Manchester \& Taylor 1981), but these attempts 
were unsuccessful.
In the two years that followed, strong evidence was accumulated to identify 
the X-ray source 1E0630+178 as the counterpart of Geminga (Bignami
et al. 1983, and references therein), and an optical counterpart 
was also detected (Bignami et al. 1988; Halpern \& Tytler 1988, and 
references therein).

The first evidence for the Geminga source being a pulsar 
came from a ROSAT observation (Halpern \& Holt 1992). 
A clear periodicity of about 237 milliseconds 
was detected, and was subsequently verified by EGRET (Bertsch et al. 
1992).
With the added advantage of the knowledge of an accurate period and its time
derivative (derived from the high energy observations), a series of radio 
observations were made hoping to detect the periodicity, but with no 
success. The apparent `radio-quiet' nature of the Geminga pulsar led to 
the suggestion that the radio emission beam is probably not directed 
towards the Earth.
Alternative suggestions were also put forward, for example, that the absense 
of radio pulses from Geminga may be due to a genuine lack of radio 
emission (Halpern \& Ruderman 1993).

After this series of unsuccessful attempts the radio investigations had 
virtually ceased during the past few years, until the 
reported detections of radio pulses from Geminga
at 102 MHz at the Pushchino Radio Observatory (Kuzmin \& Losovsky
1997; Malofeev \& Malov 1997; Shitov \& Pugachev 1997). All the three
groups reported the detection of the radio signal at roughly the same
dispersion measure, but different flux densities and pulse widths. While
Kuzmin \& Losovsky reported an average flux of about 100 mJy, Malofeev \&
Malov, and Shitov \& Pugachev found average fluxes of $60\pm 95$ mJy
and $8\pm 3$ mJy, respectively.
From their observations in December 1996 and January 1997, 
Malofeev \& Malov (1997) concluded that the observed intensity
was highly variable.

In this paper, we report the results of our observations targeted to
detect the Geminga pulsar at 35 and 327 MHz using the  
Gauribidanur Telescope and the Ooty Radio Telescope respectively.
In the following section (2), we  describe our observational
set up at the two telescopes, the analysis procedures used  and the results from
these observations.
At the end, in section
\ref{sec-discuss}, we discuss the possible reasons for the apparent
lack of detectable
radio emission from this pulsar at frequencies other than around 100 MHz.

\section{Observations, Analysis and Results}
\subsection{The 35 MHz observations using the Gauribidanur Telescope}
\label{sec-obser35}
These low-radio-frequency  observations were made using the 
Gauribidanur Telescope located at a latitude of 13$^{\circ}$.6 N in south India.
The East-West arm of the dipole T-array, with an effective collecting 
area of about 16,000 m$^2$, was used with its limited tracking 
facility (Deshpande et al. 1989). In the receiver set-up used
(Deshpande et al. 1998), 
a 1-MHz band centred at 34.5 MHz was down-converted 
to a 10-MHz IF band, which was sampled using harmonic sampling with a 
clock rate of 2.1 MHz. The signal voltages were digitized to two-bit 
4-level samples and directly recorded using a PC-based
data acquisition system. The recorded data were later processed 
off-line on a general-purpose computer. 

In the off-line processing, the data were Fourier transformed in suitable
blocks of the 2-bit samples to obtain an equivalent spectrometer output for
a chosen number of  narrow channels spanning the 1.05 MHz band, so as to allow
for suitable dedispersion. At 35 MHz, for a DM of $\sim$3 pc cm$^{-3}$,
the dispersion delay across the band is about half a second, requiring
sufficiently narrow channel widths.  For the 
Geminga data, the number of such channels was 256 (each about 4 kHz wide).
In each channel, the power output was  integrated in time 
for about 2 milliseconds. The series of
such spectra in each observation were examined for possible interference 
and where found, the corresponding time and frequencies were noted
for data rejection during futher processing steps. The interference-free
data were then folded to obtain an average profile (for each channel) over
the apparent rotation period of the pulsar (or over stretches that are
integral multiples of the period).

Such average profiles for the 256 spectral channels from a given data set
were combined after
dedispersing for a range of assumed values of the dispersion measure.
The combined profiles, one for each of the DM values, were tested for existence
of a significant `pulse' feature. The DM range of 0-60 pc cm$^{-3}$ was covered 
with finely spaced steps.
The data sets subjected to the above mentioned 
DM search test were of three types, where the average profiles in each 
of the 256
channels were obtained with a folding interval equal to  ($i$) the pulsar
period, ($ii$) twice the period  and ($iii$) ten times the period.
The case ($i$) gives the so called  average profile, which is tested for
a single pulse-feature with a significant contrast referenced to the `off-pulse' 
baseline.
The data in the case ($ii$) was tested for two similar `pulse' features
with a separation equal to the pulsar period.
In low-radio-frequency observations (like ours) where the interference 
could be severe, tests like ($ii$) are quite effective in distinguishing
between the real pulsar signal and possible artefacts due to interference
or random noise.
In these two cases, the profiles smoothed to a variety of coarser resolutions
(allowing optimum detection for a range of pulse-widths) were also tested
for `pulse' features.
The 10-period average profile (case $iii$) in each trial, was Fourier 
transformed and the spectrum was examined for a significant feature
corresponding to the rotation frequency of the pulsar.

\begin{figure}
\epsfig{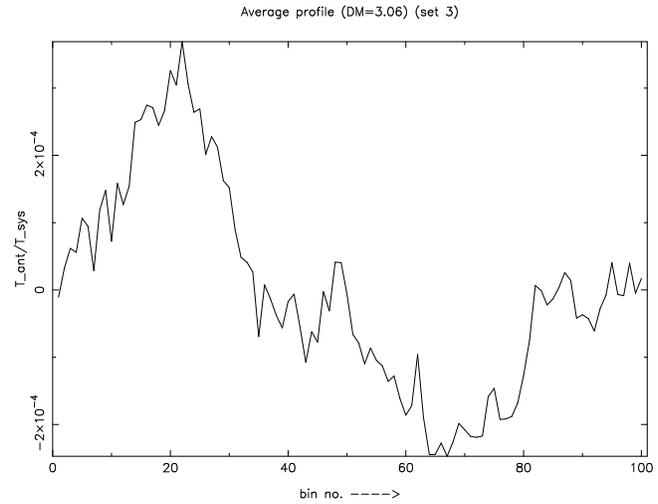}
\caption[]{Integrated profile (across one period) corresponding to set No.3
at 35 MHz, after 20-bin smoothing.}
\label{fig:set3twop}
\end{figure}

A total of seven sets available from our observations during
March-July 1997 were analysed and tested as described above.
The results of the above mentioned tests were also examined for
only  parts of the band in some cases, to assess if the main 
contribution came from only a few spectral channels. 
Tests were also conducted on sets obtained by combining two or more of the
available original sets/subsets. 
Before combining, the
cross correlation between the sets, i.e. the 2-dimensional intensity maps 
(intensity as a function of the pulse phase and frequency), was performed 
to find the relative phase shift corresponding to the maximum 
correlation. The two input sets were summed after compensating 
for the relative shift. For further combinations, the above procedure was
repeated. The advantage in such a procedure is that it looks
for the best match between the possible patterns in the two component sets
due to dispersion delay 
gradients, if any, but without
committing the process to any particular dispersion measure.
Thus, the flexibility for a DM value search is not sacrificed during
the combining procedure. 

Out of all our observations, what we consider as the `best-case' results 
came from  data sets No. 3 and 7. The set No.3, with $\sim$600 
seconds of observation, 
was the lone set that showed a somewhat significant feature at 4.22 Hz in 
the power spectrum
of the average profile over 10-periods (case $iii$). 
The feature was most prominent for DM $\sim$3 pc cm$^{-3}$. 
Also, the corresponding average profile over a two-period stretch
{\it did} show two features separated 
by one pulse period. However, it also showed a comparable additional
bump next to only one of the features in this two period stretch.  
We show in Fig. \ref{fig:set3twop} the corresponding average profile
over one-period stretch. The {\it mean} has been subtracted from 
the profile, which has been smoothed by a 20-bin window. The expected noise 
is $1.2\times 10^{-4}$ ($1\; \sigma$) in the same units as on the Y-axis. 
Although, this is one of the  `best-case' profiles we have (particularly 
as the observation was not affected by any noticeable interference), we consider 
this profile consistent with what may simply be system
noise.

\begin{figure}
\epsfig{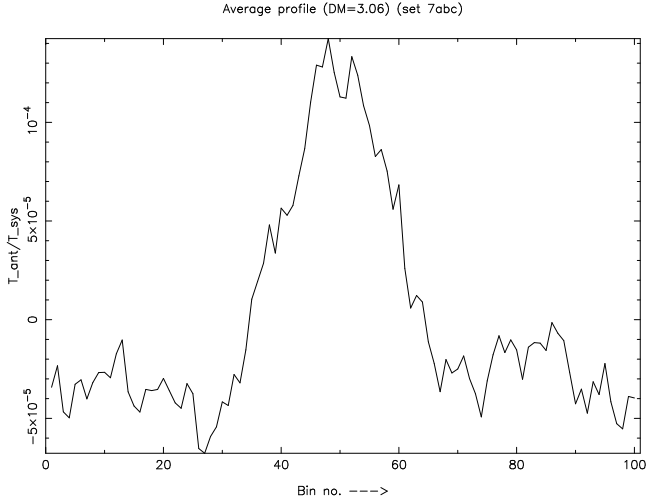}
\caption[]{Integrated profile (across one period) corresponding to set No.7
at 35 MHz, after 20-bin smoothing.}
\label{fig:set7}
\end{figure}

Another set of our observations, set No. 7, showed some interesting results. 
It had four times more integration than set No.3, but we had 
to reject many (about 20\%) frequency channels
which were affected by interference. 
Over the range of DMs examined, the power spectrum analysis of this data 
set {\it did not} show any significant feature 
at 4.22 Hz, nor did it have two similar looking pulse-like features when the 
time series was folded at twice the rotation period of the pulsar. 
However, when the data were folded at the pulse period,  and 
dedispersed (for a DM $\sim$3 pc cm$^{-3}$), the average profile showed
a significant looking pulse feature, Fig. \ref{fig:set7}.
The noise level in this profile is about $6\times 10^{-5}$ 
($1\;\sigma$), and we have used this profile to derive a useful upper
limit (3-$\sigma$) of about 75 to 100 mJy for the average flux at 35 MHz. 
We 
conclude that this profile is also entirely due to noise. In fact, 
we wish to emphasize  these  as instructive examples of how noise 
can produce very {\it appealing profiles}! It is worth mentioning that
similar situations occured at some other values of DM too, signalling
no significant preference for a DM close to 3 pc cm$^{-3}$.

\subsection{The 327 MHz observations using the Ooty Telescope}
\label{sec-obser327}
Observations at 327 MHz were conducted on 22$^{\rm nd}$ November 
1995 using the Ooty Radio Telescope (ORT), India (Swarup et al. 1971).
The telescope, an offset-parabolic cylindrical dish, has 
an effective collecting area of about 8000 
m$^2$, which corresponds to $\sim$3 K/Jy. 
With the improved feed system (Selvanayagam et al. 1993) 
the system temperature is typically about 200 K ($T_{\rm rec} + T_{\rm sky}$).
The telescope operates at a fixed centre frequency of 327 MHz (with
about 10 MHz bandwidth). As the dipoles in the feed array are
oriented North-South, the telescope is
not sensitive to the other (East-West) component of polarisation. 

The front-end electronics down-converts the signal to an intermediate
frequency (IF) of 30 MHz. This IF-signal is used as an input
to a special-purpose pulsar processor (see McConnell et al. 1996 for details),
which uses four
filters with width 2.5 MHz each, spanning the band from 25 to 35 MHz. The
signals from each of the filters are sampled using harmonic complex
sampling at the Nyquist rate. Each of these digital sample trains is Fourier
transformed to obtain 256 spectral channels across each of the 2.5 MHz bands.
Thus, each of the 1024 channels produced has a width of $\sim$10 kHz. The
total-power outputs from the channels are combined on-line after dedispersion
(for an assumed DM value) to produce a single time series with samples at
intervals of $102.4\;{\rm \mu sec}$. 
These time series were averaged synchronously with the apparent pulse 
period over blocks of every $2^{20}$ samples  separately to obtain an 
average profile across 64 bins spanning the predicted (apparent) period of the
Geminga pulsar. 
The mean total-power in each of these blocks was 
computed and subtracted from the average profile for that block. The 
subtracted means of all the blocks were stored separately and used 
as measures of the system temperature during flux calibration. 


In the present case, a  DM of 2 pc cm$^{-3}$ was used for the on-line
dedispersion. Although, the {\it true} DM may be somewhat different 
from this value,
the possible smearing due to the difference is unlikely to be serious 
becasue, 
(a) a DM of 1 pc cm$^{-3}$ corresponds to a differential
delay of only $\sim$2 ms across the band of 10 MHz at 327 MHz;
(b) the distance to Geminga is about 100 pc; and (c) the pulse duty cycle
is expected to be 10-20\%.

\begin{figure}
\epsfig{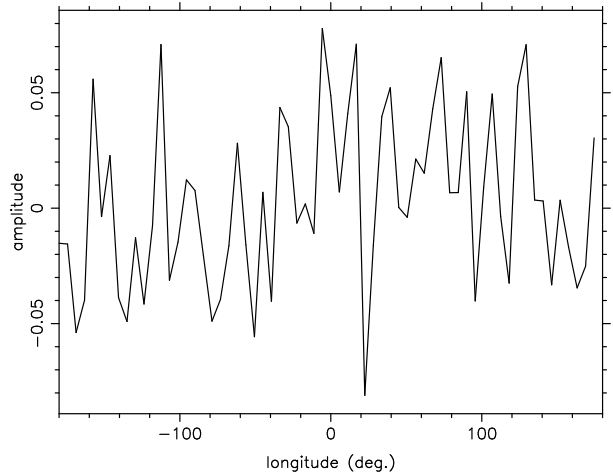}
\caption[]{Integrated profile over one-period stretch at 327 MHz.}
\label{fig:prof327}
\end{figure}

Figure \ref{fig:prof327} shows the average profile based on 
6 Hrs of integration during our 327 MHz observation towards the Geminga pulsar.
As can be seen, there is no pulsed signal above our
detection limit.
If we assume that the duty 
cycle of the radio counterpart of Geminga is the same as that observed 
in $\gamma$--rays ($\sim 20$\%), then our 327 MHz observation implies
an upper limit of 0.2--0.3 mJy for the average flux (corresponding to
a $3\sigma$ level).
To the best of our knowledge, this represents  
by far the most stringent limit available in the frequency range 300-400 MHz.

 The uncertainty in the quoted limits is mainly due to the corresponding
uncertainty in the flux calibration. At both the frequencies, the telescope
gins were calibrated using observations on continuum sources.
  
\section{Discussions}
\label{sec-discuss}
We have reported here the results of our observations aimed at 
searching for the radio 
counterpart of the Geminga pulsar at 327 MHz and 35 MHz. 
At both these frequencies, we did not find any significant pulsed emission
allowing us to estimate useful upper limits,
of 75--100 mJy at 35 MHz and 0.2--0.3 mJy at 327 MHz, for 
the average pulsed emission at the two well separated frequencies.
In the light of these and many other earlier attempts, 
the reported detection of this pulsar at 102 MHz 
by Malofeev \& Malov (1997) appears puzzling. However, as they themselves report, 
the strongest support for their positive detection of the pulsar at 102 MHz 
comes from the fact that the pulses on each observing session arrived at 
the expected phase pre-calculated on the basis of the available ephemeris. 

They also observed that  the pulsing flux at 102 MHz is highly 
variable ($S_{\rm 102} = 60\pm 95$ mJy). With the mean value of 60 mJy and 
with a spectral index of $\sim -2$, we would expect to receive an average
flux of more than about 
6 mJy from this pulsar at 327 MHz.  However, as mentioned above, we do not 
see any significant emission  above a 0.2 mJy level. This would imply
a spectral index for the pulsed emission to be $<-4.5$, giving this 
pulsar the steepest spectrum in this frequency range.
The observed intensities of nearby pulsars, like J0437--4715,
 are known to fluctuate by a factor of a few due to interstellar 
diffractive scintillation, with time scales of the order of an hour or so. 
We observed Geminga at 327 MHz for about 6 hours, and we consider it 
unlikely that the pulsar would have stayed at  a `low-state' due to 
scintillations during most of that observation. 

From the properties of about 800 known pulsars, we expect the pulsed emission
from a pulsar to be detectable over a reasonably wide range of radio 
frequencies. However, given the claimed detection at 102 MHz, and our
failures at 35 and 327 MHz, we are left to speculate on the possibilities of 
having very weak or no emission at frequencies other than around 102 MHz.
One possibility, where the 102 and 327 MHz results would appear consistent
is that the trajectory of our line-of-sight is at the edge of the emission 
cone at 102 MHz. Then, at higher frequencies, the line-of-sight is more 
likely to  miss the emission cone, if it becomes smaller with increasing 
frequency. This should not be too surprising, as we have at least one known 
example, PSR B0943+10, where this is almost certainly the case. The observed 
flux from this pulsar drops down quite rapidly at frequencies above roughly 
600 MHz (Malofeev \& Malov 1980; Lorimer et al. 1995) while on the 
other hand, it does not show a spectral turn-over down to 35 MHz (Deshpande 
\& Radhakrishnan 1992; 1994). The non-detection of the Geminga pulsar at 35 
MHz can also be understood in such a case, if the spectrum has a turnover in 
the range 35-100 MHz. From observations of other pulsars, we know that such 
spectral turn-overs at about 60-100 MHz are not unusual (see Deshpande \& 
Radhakrishnan 1992, for example).

The other possibility is that the reported value of the average flux at 102 
MHz (Malofeev \& Malov 1997; Kuzmin \& Losovsky 1997) is an over-estimate and 
the value is below or about 10 mJy, consistent with that found by Shitov \& 
Pugachev (1997). In such a case, a straight spectrum from 35 to 300 MHz, 
implying a spectral index of about --2.5, would be consistent with our limits.
\vspace*{0.5cm}

\noindent
To conclude, the main points are summarised below.

\vspace*{0.25cm}

\begin{itemize}
\item We have searched for pulsed radio emission from Geminga at 35 and 327 MHz,
      using a variety of stringent detection criteria.
\item At both the frequencies, we {\it did not detect} any significant emission 
      from the source, 
      implying upper limits for the average flux; namely, 
      75--100 mJy at 35 MHz, and 0.2--0.3 mJy at 327 MHz.
\item We suggest that the lack of emission from the pulsar at 
      frequencies other than around 102 MHz may be understood in terms of
      a peculiar geometrical situation or in terms of a large error in the
      estimated average strength at 102 MHz.
\end{itemize}

\begin{acknowledgement}
We would like to thank Ashish Asgekar for his help during the 35 MHz 
observations, V. Radhakrishnan for his encouragement and many useful
discussions during this work, and the referee R. Wielebinski for his comments
on the manuscript. We thank J. G. Ables and D. McConnell for making their 
pulsar processor available for our observations at 327 MHz.
\end{acknowledgement}

\end{document}